\documentclass[preprint, double-spaced]{revtex4}
\usepackage{graphicx}
\usepackage{amsmath,bbm}
\usepackage{amssymb,bm}
\usepackage{array}
\begin{document}
\title{Unified description of charmonium suppression in a quark-gluon plasma medium at RHIC and LHC Energies}
\author{Captain R. Singh$^{1}$}
\author{P. K. Srivastava$^{2}$\footnote{prasu111@gmail.com}}
\author{S. Ganesh$^{1}$}
\author{M. Mishra$^{1}$\footnote{madhukar.12@gmail.com}}
\affiliation{$^1$ Department of Physics, Birla Institute of Technology and Science Pilani, Pilani - 333031, INDIA}
\affiliation{$^2$ Department of Physics, Indian Institute of Technology 
Roorkee, Roorkee - 247667, INDIA}
\begin{abstract}
Recent experimental and theoretical studies suggest that the quarkonia suppression in a thermal QCD medium created at heavy ion collisions is a complex interplay of various physical processes. In this article we put together most of these processes in a unified way to calculate the charmonium survival probability (nuclear modification factor) at energies available at relativistic heavy ion collider (RHIC) and large hadron collider (LHC) experiments. We have included shadowing as the dominant cold nuclear matter (CNM) effect. Further, gluo-dissociation and collision damping have been included which provide width to the spectral function of charmonia in a thermal medium and cause the dissociation of charmonium along with usual colour screening. We include the colour screening using our recently proposed modified Chu and Matsui model. Furthermore, we incorporate the recombination of uncorrelated charm and anti-charm quark for the regeneration of charmonium over the entire temporal evolution of QGP medium. Finally we do the feed-down correction from the excited states to calculate the survival probability of charmonium. We find that our unified model suitably describes the experimental nuclear modification data of $J/\psi$ at RHIC and LHC simultaneously.
\\

\end{abstract}

\maketitle 
\section{Introduction}
\noindent
The physical picture of quarkonium dissociation in a thermal medium has undergone theoretical and experimental refinements over the last decade~\cite{QWG}. Heavy quarkonia ($J/\psi,~\Upsilon$ etc.) suppression is considered as the most classical observable of QGP formation in heavy ion collision experiments. This is because the heavy mass scale ($m=3.1$ GeV  for $J/\psi$ and $m=9.2$ GeV for $\Upsilon$) makes these system possible for analytical treatment theoretically. On the other side, decay of heavy quarkonia via dileptonic channel lead to relatively clean signal which can be precisely measured experimentally. Till the mid-2000s, Debye screening was thought to be the only possible mechanism for the anomalous suppression of charmonium ($J/\psi)$ and bottomonium ($\Upsilon$)~\cite{matsui} in QGP medium. However, experimental results involve some puzzling features which defy explanations based on color screening alone~\cite{abreu,arnaldi,adare,cms,alice}. The first such experimental result is the less suppression at mid-rapidity than forward rapidity observed at RHIC and also at LHC~\cite{cms,alice} which is in contradiction to the color screening scenario (as color screening predicts larger suppression at higher density region of plasma which is actually the mid-rapidity). Second such experimental result is the same amount of charmonium suppression at SPS and RHIC energies for the same number of participants~\cite{arnaldi}. Although, the available energy spans over two order of magnitude in moving from SPS to LHC, the amount of charmonium suppression is found to be similar. Regeneration of charmonia in QGP through the recombination of $c$ and $\bar{c}$ quarks is believed as the main reason for this experimental observation. Third experimental observation is the suppression pattern in forward and backward rapidity in $d-Au$ collision at RHIC. A suppression is observed at forward rapidity (in the d-going direction) and an enhancement at backward rapidity (in the Au-going direction)~\cite{phenix}. This result suggests the importance of charmonia break-up effects in nuclear matter at final stages of collision apart from usual cold-nuclear-matter effects in the initial stage. All these experimental observations suggest that the charmonium suppression in QCD plasma is not the result of a single mechanism, but is a complex interplay of various physical processes.\\ 

On the theoretical side, the development in thermal field theory shows that the static potential between two heavy quarks placed in a QCD medium consists of two parts~\cite{laine,lata}. It is the first part which represents the standard time-independent Debye-screened potential (Earlier it was thought that the Debye-screened potential is only the dominant part in heavy-quark potential based on which one can understand the dissociation of quarkonia in QGP). The second part of potential, other than the standard Debye-screened part, is imaginary and in the limit of $t\rightarrow\infty$, represents the thermal decay width induced by Landau damping of the low frequency gauge fields that mediate interaction between two heavy quarks~\cite{laine}. Later Brambilla et al.~\cite{brambilla} have shown that the apart from thermal width originates from the imaginary part of the gluon self energy, singlet to octet transition of heavy meson resonance due to the gluonic interaction also contribute to the decay width of quarkonia states. In a QGP, gluons can collide with a color-singlet heavy quarkonium leading to its dissociation~\cite{wong}. Dissociation by the absorption of a single gluon is allowed as the color-octet final state of a free quark and anti-quark can propagate in the colored QGP medium, in contrast to the color-less hadronic medium. One of the earlier treatment of the dissociation of heavy quarkonium by the absorption of a $E1$ gluon (where $E1$ is the lowest electric mode for the spin-orbital wave function of gluons) was carried out by Peskin and Bhanot~\cite{peskin,bhanot}. Therefore, the dissociation of quarkonia does not happen only due to the Debye screening but it can occur by the gluonic dissociation and the collisional damping (Landau damping) as well.\\ 

The production of charmonium in nucleon-nucleon collisions is also an involved process~\cite{bodwin1}. The charmonia production process in elementary hadronic collision e.g., $p+p$ collision, begins with the formation of a $c\bar{c}$ pair; this pair can then either lead to open charm production or subsequently bind to form a charmonium state (about $10\%$ for all charmonia)~\cite{satz}. The dominant high energy production mechanism for charmoina is gluon-gluon fusion. Based on the scales involved, the production process of charmonia is believed to be factorisable into two parts: a charm and anti-charm quark produced through nucleon-nucleon collision is a perturbative QCD process~\cite{bodwin1}. However, the formation and evolution of this pair into a meson is governed by non-perturbative QCD. Hence, heavy-quarkonia provide a unique laboratory which can explore the interplay of perturbative and non-perturbative QCD effects. A variety of theoretical approaches has been proposed in the literature to calculate the heavy quarkonium production in nucleon-nucleon collisions~\cite{bodwin2,bodwin3,kang1,kang2,baranov1,baranov2}. Non-relativistic QCD (NRQCD)~\cite{bodwin2,bodwin3} and fragmentation approaches~\cite{kang1,kang2} are the two theoretical methods based on QCD which are being used in most of the quarkonium production and suppression models. However, the heavy quarkonium production mechanism is still a topic of intense debate. \\

The heavy quarkonia production processes, in the case of nucleus-nucleus collision, are significantly affected by the nuclear environment~\cite{zhou}. These effects are known as cold nuclear matter (CNM) effects. In most of the literature, three CNM effects are considered. The first and dominant CNM effect in the case of quarkonium production is shadowing. Change in the parton distribution function in the nucleus which control the initial parton behaviour and strongly depends on the collisional kinematics ( in the small $x$ region nuclear parton distribution function is clearly suppressed compared to that of nucleon) is known as shadowing~\cite{muller}. The shadowing cause the production cross-section to become less in $A-A$ case to that of pure N-N collision. Second CNM contribution is known as Cronin effect~\cite{cronin,hufner}. It describes the initial gluon multi-scattering with the neighbouring nucleons presented in the nucleus prior to hard scattering and the quarkonia formation. This results in the broadening of transverse momentum distribution of produced charmonia. Nuclear absorption~\cite{gerschel} is another CNM contribution to the charmonia production. The interaction between charmonium and the primary nucleons leads to the normal suppression of charmonia. Nuclear absorption is the dominant CNM effect at lower energies. However, the cross-section for nuclear absorption decreases with the increase in energy~\cite{lourenco}.\\

Recently, it has been proposed that recombination of initially uncorrelated $c$ and $\bar{c}$ quarks in QGP can also regenerate the charmonia states~\cite{pbm,andronic,rapp1,rapp2,thews1,thews2,thews3}. The calculation of regeneration of charmonium is based on statistical hadronization model~\cite{pbm,andronic} and kinetic model in which the $J/\psi$ production is described via dynamical melting and regeneration over the whole temporal evolution of the QGP~\cite{rapp1,rapp2,thews1,thews3}. Some transport calculations were also performed to calculate the number of regenerated $J/\psi$s~\cite{zhang,cassing}. At lower energies, this contribution is very low almost negligible because of the fewer number of initially produced charm quarks. However, at higher energies of RHIC and LHC, the regeneration factor become important. Thus $J/\psi$ whose suppression is actually proposed as a signal to confirm the existence of quark gluon plasma earlier can also turn out to provide extremely useful probe for QCD medium created at heavy ion collision experiments.\\

In this article we present a unified model which includes most of the above discussed dissociation as well as production (recombination) processes to finally calculate the survival probability of $J/\psi$ in QGP medium. We have constructed this model based on the kinetic approach whose original ingredients was given by Thews et al.~\cite{thews1,thews2,thews3}. In this approach, there are two terms written on the basis of Boltzmann kinetic equation. First term, which we call as dissociation term, includes the dissociation processes like gluonic dissociation and collisional damping. The second term (formation term) provides the (re)generation of $J/\psi$ due to the recombination of charm-anticharm quark. These two terms compete over the entire temporal evolution of the QGP and at freeze-out temperature we get the multiplicity of finally survived $J/\psi$s. To include the gluonic dissociation we took help of a model which was developed by Nendzig and Wolschin~\cite{nendzig} and later used by two of the authors~\cite{ganesh}. The thermal width due to collisional damping is calculated based on the thermal field theory as discussed by M. Laine et. al.~\cite{laine}. We have also included the shadowing effect to incorporate the CNM effect properly into the production process. We consider color screening as a dissociation process of charmonium acting till the formation of charmonium bound states followed by gluonic dissociation along with the collisional damping. It means that the color-screening is active at the initial times of medium evolution when the temperature is high enough to melt-down the charmonia states. Later the dissociation probability by color screening diminishes rapidly and becomes zero at lower temperatures (time larger than the quarkonia formation time). To include the dissociation of $J/\psi$ due to Debye screening (color screening) in the QCD plasma, we have used a new model constructed by two of the authors based on the color screening in the QGP~\cite{pks1,pks2}. In this color screening model we have used the quasipaticle (QPM) equation of state (EOS) to describe the basic partonic properties of QGP phase. To define the dynamics of the system created in the heavy ion collisions, we have used the (1+1)-dimensional viscous hydrodynamics. We have included only the shear viscosity and neglect the bulk viscosity. We have also suitably incorporated the overall feed-down correction from the higher charmonium states ($\chi_{c}$ and $\psi^{'}$) to the charmonium ($J/\psi)$. 
\section{Model Formalism}
The abundance of charm quark, anti-quark and their bound state i.e, charmonia states ($J/\psi,~\chi_{c},~\psi^{'}$ etc.) is governed by a simple master equation involving two reactions: the formation reaction and the dissociation reaction. Thus the time evolution of the number of bound charmonium state in the deconfined region can be written as~\cite{thews1}:
\begin{equation}
 \frac{d N_{J/\psi}}{d\tau} = \Gamma_{F,nl} N_{c}~N_{\bar{c}}~[V(\tau)]^{-1} -
\Gamma_{D,nl} N_{J/\psi}.
\label{tq}
\end{equation}
In the above equation, the first term in the right hand side represents the formation term by recombination of uncorrelated charm quark and anti-quark. The second term in the right hand side is the dissociation term of charmonium. $\Gamma_{D,nl}$ and $\Gamma_{F,nl}$ are the dissociation width and recombination reactivity corresponding to the dissociation and regeneration of charmonia, respectively. It is important here to mention that the unit of $\Gamma_{D,nl}$ is in GeV or $fm^{-1}$. However, the recombination reactivity $\Gamma_{F,nl}$ has its units in $fm^{2}$ or GeV$^{-2}$. It only change in the unit of $fm^{-1}$ or $GeV$ when it is multiplied by the inverse of the system volume $[V(\tau)]^{-1}$. $N_{c}$, $N_{\bar{c}}$ and $N_{J/\psi}$ are the numbers of produced charm, anti-charm and $J/\psi$, respectively. At the initial time, we have taken $N_{c}$=$N_{\bar{c}}$=$N_{c\bar{c}}$. When the total number of regenerated charmonia is less than the initial number of $N_{c\bar{c}}$, one can obtain the analytical solution of Eq.(\ref{tq}) as follows~\cite{thews2}: 
\begin{equation}
 N_{J/\psi}(\tau_{f},b) = \epsilon(\tau_{f}) \left[ N_{J/\psi}(\tau_{0},b) +
N_{c\bar{c}}^{2} \int_{\tau_{0}}^{\tau_{f}} \Gamma_{F,nl} [V(\tau)
\epsilon(\tau)]^{-1} d\tau \right],
\label{tq1}
\end{equation}
where $\tau_{0}$ is the initial thermalization time of the QGP and $\tau_f$ is the life time of the QGP. $N_{J/\psi} (\tau_{0}, b)$ is the initial multiplicity and $N_{J/\psi} (\tau_{f}, b)$  is the finally survived number of $J/\psi$ meson. The variables used in the Eq. (\ref{tq1}),  $\epsilon(\tau_{f})$ and $\epsilon(\tau)$ are the suppression factors which can be obtained using the following expressions :
\begin{equation}
\epsilon(\tau_{f})= \exp{\left[-\int_{\tau_{0}}^{{\tau_{f}}} \Gamma_{D,nl} \;d \tau\right]},
\end{equation}
and 
\begin{equation}
\epsilon(\tau) = \exp{\left[-\int_{\tau_{0}}^{{\tau}}\Gamma_{D,nl} \; d \tau\right]}.
\end{equation}
We have defined $\Gamma_{D,nl}$ as the net sum of collisional damping reaction rate ($\Gamma_{damp,nl}$) and gluonic dissociation reaction rate ($\Gamma_{gd,nl}$) of charmonia in QGP, given as; 
\begin{equation}
 \Gamma_{D,nl} = \Gamma_{gd,nl} + \Gamma_{damp,nl}.
\label{netd}
\end{equation}
The initial time ($\tau_{0}$) is taken as the formation time required for the charmonia formation and where the dissociation due to color screening becomes zero. $\tau_{0}$ is taken as $0.89,~2.0$ and $1.5$ for $J/\psi$, $\chi_{c}$ and $\psi^{'}$, respectively~\cite{helmut}. 
\subsection{Gluonic Dissociation and Collisional Damping}
The dissociation of $c\bar{c}$ bound state due to gluonic dissociation along
with collisional damping in QGP medium was formulated by Wolschin et el.~\cite{nendzig,wong}.
Here we briefly discussed about these dissociation mechanisms.\\
\vspace{2mm}
\subsubsection{Collisional Damping}
To determine the collisional dissociation and gluonic dissociation rate, we take the help of effective potential models. In our work, we have used the singlet potential for  $c-\bar{c}$ bound state in the QGP as follows~\cite{nendzig};
\begin{eqnarray}
 V(r,m_D) = \frac{\sigma}{m_D}(1 - e^{-m_D\,r}) - \alpha_{eff} \left ( m_D
+ \frac{e^{-m_D\,r}}{r} \right ) \nonumber \\-  i\alpha_{eff} T \int_0^\infty
\frac{2\,z\,dz}{(1+z^2)^2} \left ( 1 - \frac{\sin(m_D\,r\,z)}{m_D\,r\,z} \right),
\end{eqnarray}
where the first and second term in the right hand side is the string term and the columbic term, respectively. The third term in the right hand side is the imaginary part of the heavy-quark potential. In the above Eq. (6) :  
\begin{itemize}
 \item $\sigma$ is the string tension constant between $c\bar{c}$ bound
state, given as  $\sigma= 0.192$ GeV$^2$.
\item $m_{D}$ is Debye mass, $m_D = T\sqrt{4\pi \alpha_s^T \left (
\frac{N_c}{3} + \frac{N_f}{6} \right ) }$, and $\alpha_s^T$ is coupling
constant at hard scale, as it should be $\alpha_{s}^{T} = \alpha_{s}(2\pi
T)\leq 0.50$~\cite{nendzig}. For charmonia we found $\alpha_s^T \simeq 0.25$ .
Here we have considered, $N_{c} = 3$ \& $N_{f} = 3$ and evolution of
temperature, $T$ as the function of time, $\tau$ and impact parameter, $b$~
\cite{ganesh}.
\item $\alpha_{eff}$ is effective coupling constant, depending on the strong
coupling constant at soft scale $\alpha_{s}^{s} =
\alpha_{s} (m_{c} \alpha_{s}/2) = 0.4725$, given as  $\alpha_{eff} =
\frac{4}{3}\alpha_{s}^{s}$.
\end{itemize}
 The decay rate, $\Gamma_{damp,nl}$, accounts for collisional damping by the
QGP partons. The imaginary part of potential causes the collisional damping (also termed as Landau damping in the literatures). Therefore, the decay rate can be obtained, in a first order perturbation, by folding of imaginary part of the potential with the radial wave function and is given by;
\begin{equation}
 \Gamma_{damp,nl} = \int[g_{nl}(r)^{\dagger} \left [ Im(V)\right ] g_{nl}(r)]dr,
\label{cold}
\end{equation}
here, $g_{nl}(r)$ is charmonia wave function. Corresponding to different
values of $n$ and $l$, we have obtained the wave functions for $1S (J/\psi)$,
$1P(\chi_{c})$ and $2S(\psi^{'})$  by solving the Schr\"{o}dinger equation.
\subsubsection{Gluonic Dissociation}
In a QGP, gluons can collide with a color-singlet heavy quarkonium leading to its dissociation. The ultra-soft gluon makes the color-singlet object to a color-octet object which further dissociates at the time of freeze-out. The cross-section for this gluonic dissociation process can be given as follows~\cite{nendzig}:   
\begin{equation}
\sigma_{d,nl}(E_g) = \frac{\pi^2\alpha_s^u E_g}{N_c^2} \sqrt{\frac{m_c}{E_g
+ E_{nl}}} \left(\frac{l|J_{nl}^{q,l-1}|^2 +
(l+1)|J_{nl}^{q,l+1}|^2}{2l+1} \right),
\end{equation}
where, $J_{nl}^{ql^{'}}$ is the probability density obtained using the singlet
and octet wave functions as, 
\begin{equation}
 J_{nl}^{ql'} = \int_0^\infty dr\; r\; g^*_{nl}(r)\;h_{ql'}(r)
\end{equation}
and
\begin{itemize}
\item $m_c = 1.3$ GeV, is the mass of the $c$ and $\bar{c}$.
\item $\alpha_s^u \simeq 0.59$ \cite{nendzig}, is coupling constant, scaled as
$\alpha_{s}^{u} = \alpha_{s}(\alpha_{s}m_{c}^{2}/2)$.
\item $E_{nl}$ is energy eigen values corresponding to charmonia wave
function, $g_{nl}(r)$.
\item the radial wave function $h_{ql'}(r)$ has been obtained by solving the Schr\"{o}dinger equation with the octet potential $V_{8} = \alpha_{eff}/8r$ and the value of $q$ is determined from energy conservation, $q = \sqrt{m_c(E_{g} + E_{nl})}$. 
\end{itemize}
\begin{figure}[ht]
\includegraphics[scale=0.35]{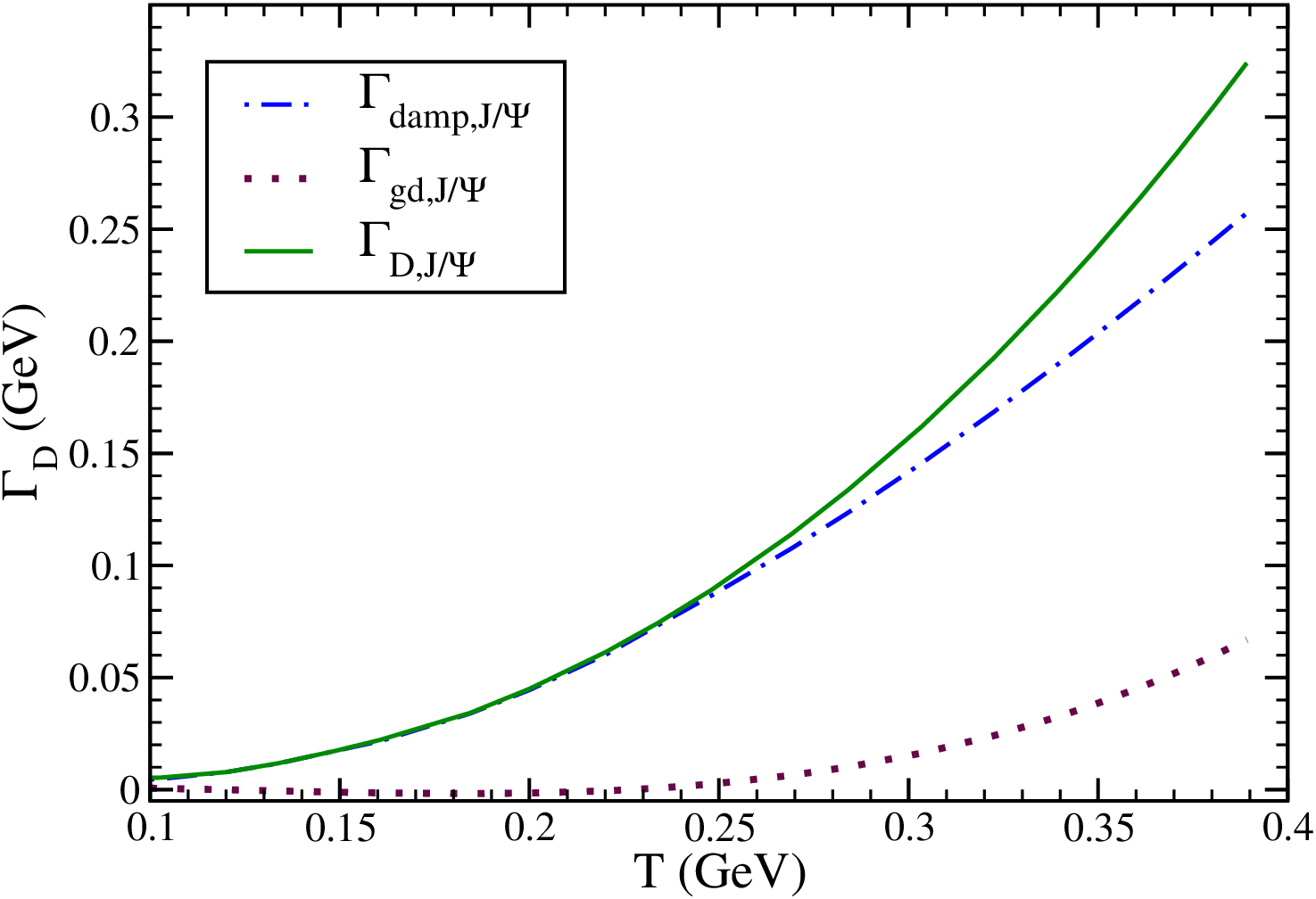}
\caption{(Color online) Variation of total dissociation width with temperature along with its components i.e., gluonic dissociation width and width due to collisional damping.\\}
\end{figure}
The Schr\"{o}dinger equation has been solved by taking a $10^{4}$ point
logarithmically spaced finite spatial grid and solving the resulting matrix
eigen value equations~\cite{ganesh}. For the octet modeling the potential is repulsive ,
which implies that the quark and antiquark can be far away from each other. To
account for this, the finite spatial grid  is taken over a very large distance,
namely $10^{2}$, as an approximation for infinity. The octet wave function
corresponding to large $c\bar{c}$ distance have negligible contribution to the
gluonic dissociation cross section.\\\\

To obtain the gluonic decay rate $\Gamma_{gd,nl}$, we have calculated the mean
of gluonic dissociation cross section by taking its thermal average over Bose-Einstein distribution function for gluons. Thus the rate of gluonic dissociation can be written as:
\begin{equation}
 \Gamma_{gd,nl} = \frac{g_d}{2\pi^2} \int_0^\infty
\frac{dp_g\,p_g^2 \sigma_{d,nl}(E_g)}{e^{E_g/T} - 1}
\label{glud}
\end{equation}
 where $g_d = 16$ is the number of gluonic degrees of freedom. The
Eq.(\ref{glud}) has been derived for idealized case where $J/\psi$ is at rest
in a thermal bath of gluons.\\\\     
\begin{figure}[ht]
\includegraphics[scale=0.35]{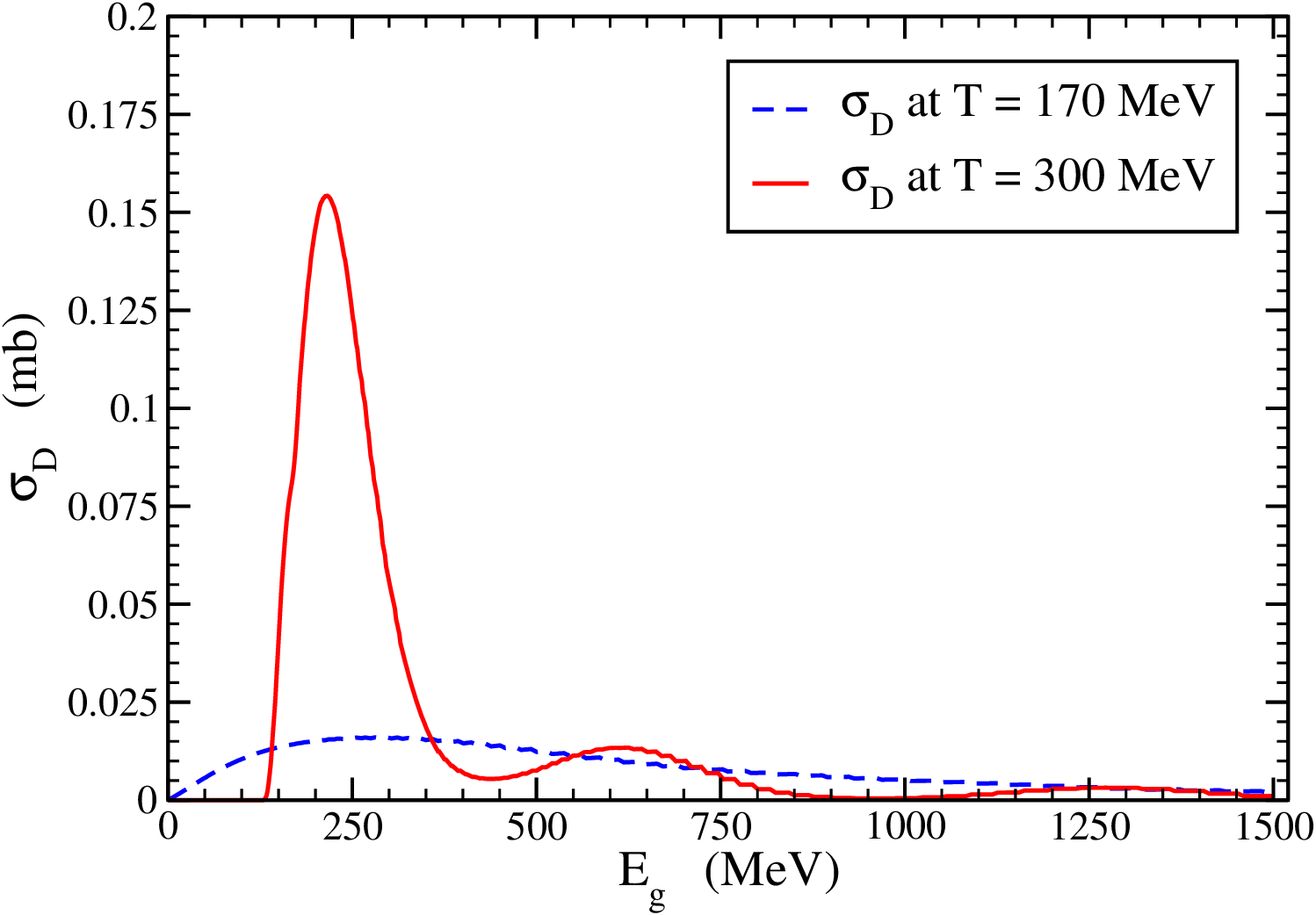}
\caption{(Color online) Variation of gluo-dissociation cross section of $J/\psi$ with respect to gluon energy, $E_g$ at $T=0.170$ GeV and $T=0.300$ GeV.\\}
\end{figure}

The total dissociation rate along with its components i.e., gluonic dissociation and collisional damping decay rates versus temperature are shown in Fig. 1. This figure depicts that the decay rate of charmonia due to gluonic dissociation remains insignificant till $T=0.250$ GeV and increases slowly beyond this temperature. Whereas, collisional damping decay rate contributes significantly throughout whole temperature range. Gluonic dissociation cross section for $J/\psi$ versus gluon energy, $E_g$ at the temperatures, $T=0.170$ GeV and $T=0.300$ GeV are also depicted in Fig. 2. 

\subsection{Regeneration via $c$ and $\bar c$ quarks}
The recombination reactivity, $\Gamma_{F,nl}$ required in Eq. (2) is calculated by taking the thermal average of product of recombination cross section and initial relative velocity between $c$ and $\bar c$, $<\sigma_{f,nl}\;v_{rel}>_{p_{c}}$ using modified Fermi-Dirac distribution function of charm quark at temperature $T$ as follows \cite{thews3};
\begin{equation}
 \Gamma_{F,nl} =
\frac{\int_{p_{c,min}}^{p_{c,max}}\int_{p_{\bar{c},min}}^{p_{\bar{c},max}}
dp_{c}\; dp_{\bar{c}}\; p_{c}^{2}\;p_{\bar{c}}^{2}\;
f_{c}\;f_{\bar{c}}\;\sigma_{f,nl}\;v_{rel}
}{\int_{p_{c,min}}^{p_{c,max}}\int_{p_{\bar{c},min}}^{p_{\bar{c},max}}
dp_{c}\; dp_{\bar{c}}\; p_{c}^{2}\;p_{\bar{c}}^{2}\;
f_{c}\;f_{\bar{c}}},
\end{equation}
where, $p_{c}$ and $p_{\bar{c}}$ are momentum of charm and anti-charm quark,
respectively. 
The $f_{c,\bar{c}}$ is the modified Fermi-Dirac distribution
function of charm and anticharm quark given as, $f_{c,\bar{c}} =
\lambda_{c,\bar{c}}/(e^{E_{c,\bar{c}}/T} + 1)$, here $E_{c,\bar{c}} =
\sqrt{p_{c,\bar{c}}^{2} + m_{c,\bar{c}}^{2}}$ is the energy of charm and
anticharm quark in medium with mass, $m_{c,\bar{c}} = 1.3$ GeV and
$\lambda_{c,\bar{c}}$ is their respective fugacity term~\cite{dks}. We have
calculated relative velocity of
$c\bar{c}$ pair in medium, given as:
\begin{equation}
v_{rel} =
\sqrt{\frac{(p_{c}^{\mu}\;
p_{\bar{c} \mu})^{2}-m_{c}^{4}}{p_{c}^{2}\;p_{\bar{c}}^{2}
+ m_{c}^{2}(p_{c}^{2} + p_{\bar{c}}^{2}  + m_{c}^{2})}}
\end{equation}
The recombination cross section $\sigma_{f,nl}$ has been obtained using
detailed balance from the dissociation cross section $\sigma_{d,nl}$
~\cite{thews1} as follows: 
\begin{equation}
 \sigma_{f,nl} = \frac{48}{36}\sigma_{d,nl}
\frac{(s-M_{nl}^{2})^{2}}{s(s-4\;m_{c}^{2})}.
\end{equation}
Here, $M_{nl} = (M_{J/\psi},\;\chi_{c},\;\psi^{'})$, is the mass of the charmonium
states. The $s=(p_1+p_2)^2$ is related to the center of mass energy of the $c\bar c$ pair with $p_1$ and $p_2$ as four momentum of $c$ and $\bar c$ quarks, respectively.\\

The $\Gamma_{F}$ for $J/\psi$ versus temperature $T$ at both, RHIC and LHC energies are shown in Fig. 3. Its value at RHIC energy is found to be larger as compared to corresponding number at LHC energy. This trend may be due to the low initial momenta of charm and anti-charm quark which participate in the J/psi formation at RHIC compared to LHC. It is also obvious from the above figure that the recombination reactivity, $\Gamma_{F}$ at LHC increases with temperature, attains maximum value at around $T=0.225$ GeV and vanishes at $T=0.300$ GeV. Whereas, at RHIC energy it peaks at $T=0.300$ GeV and remains finite even at temperature $T=0.400$ GeV.\\\\  
\begin{figure}[ht]
\includegraphics[scale=0.35]{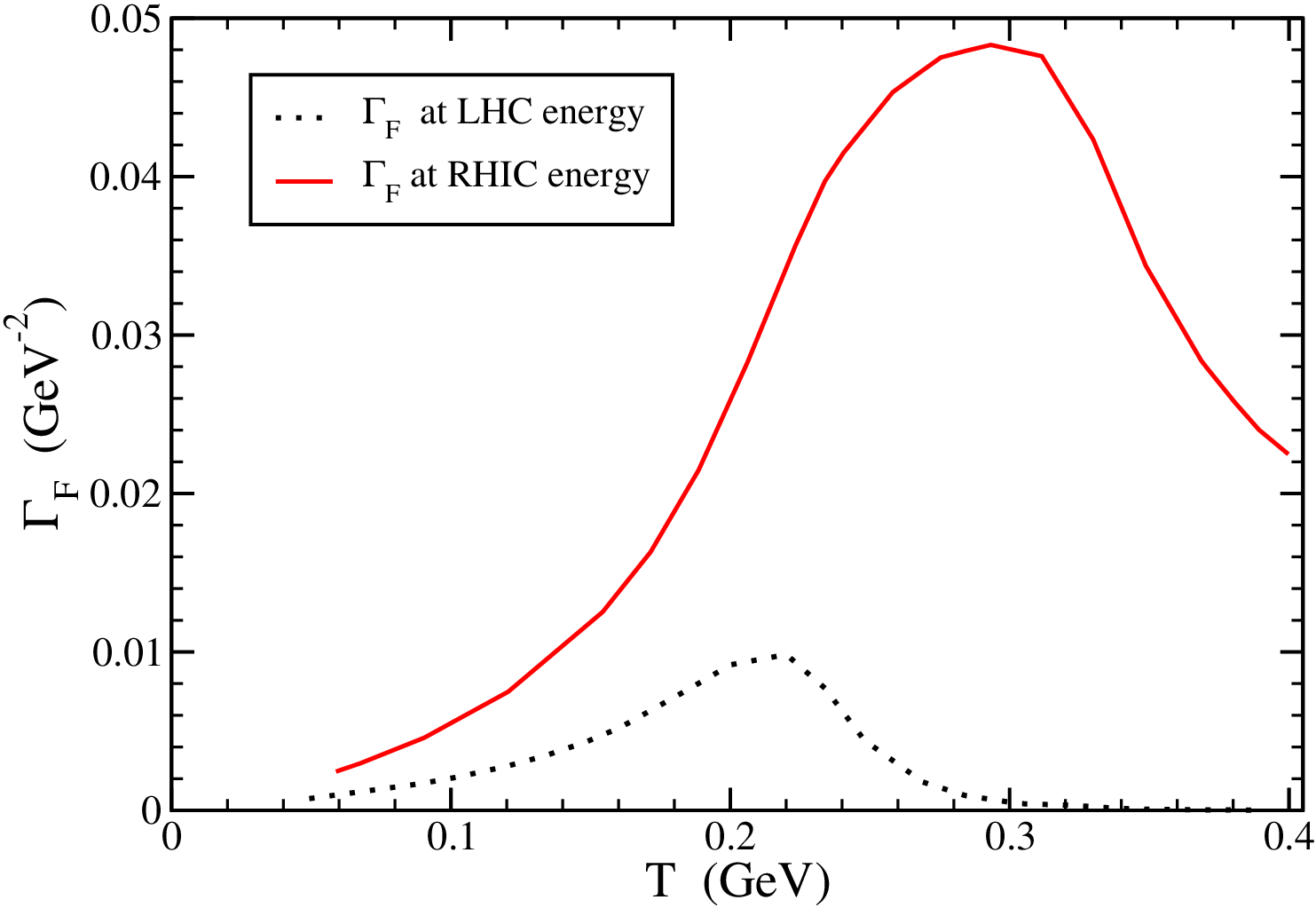}
\caption{(Color online) Variation of recombination reactivity with respect to temperature at RHIC ($\sqrt{s_{NN}}=200$ GeV) and at LHC ($\sqrt{s_{NN}}=2.76$ TeV).\\}
\end{figure}
  
\subsection{Inputs to the model}
In this section we provide the prescriptions to calculate various inputs which have been used in our model. $ N_{J/\psi}(\tau_{0})$ and $N_{c\bar{c}}$ in Eq. (2) is the initially produced $J/\psi$ and $c\bar{c}$ pair in heavy ion collisions. We have calculated these quantities using Glauber model, for per event as follows;
\begin{equation}
 N_{J/\psi}(\tau_{0}, b) = \sigma_{J/\psi}^{NN}\; T_{AA}(b)
\end{equation}
here, $T_{AA}(b)$ is nuclear overlap function, its impact parameter ($b$)
dependent values have been taken from Ref.~\cite{cern}. The 
$\sigma_{J/\psi}^{NN}$ is the cross section for production in p$+$p collision
and its values for RHIC at $\sqrt{s} = 200$ GeV and for LHC at $\sqrt{s} = 2.76$
TeV are given in Table I.
\begin{center}
\begin{tabular}{ |p{2.0cm}|p{2.0cm}|p{2.0cm}|p{2.0cm}|  }
 \hline
 \multicolumn{4}{|c|}{Table I. All cross sections are in mili-barn (mb) unit} \\
 \hline
 &$\sigma_{J/\psi}^{NN}$ & $\sigma_{\chi_{c}}^{NN}$ & $\sigma_{\psi^{'}}^{NN}$\\
 \hline
 LHC &0.0072 \cite{ts}    &1.0$\sigma_{J/\psi}^{NN}$&  
0.30$\sigma_{J/\psi}^{NN}$\\
\hline
 RHIC&   0.00139 \cite{ts}  &1.0$\sigma_{J/\psi}^{NN}$  
&0.30$\sigma_{J/\psi}^{NN}$\\
 \hline
\end{tabular}
\end{center}
Similarly, we have calculated $N_{c\bar{c}}$ from Glauber model;
\begin{equation}
 N_{c\bar{c}} (b) = \sigma_{c\bar{c}}^{NN}\; T_{AA},
\end{equation}
where $\sigma_{c\bar{c}}^{NN}$ is the cross section for $c\bar{c}$ pair
production in p$+$p collision. The $\sigma_{c\bar{c}}^{NN}$ has been calculated
using pQCD approach for GRV HO hadronic structure function \cite{thews2}, we
have obtained $\sigma_{c\bar{c}}^{NN} = 3.546$ mb for LHC at $\sqrt{s} = 2.76$
TeV and  $\sigma_{c\bar{c}}^{NN} = 0.346$ mb for RHIC at  $\sqrt{s} = 200$ GeV.
The quantity $V(\tau)$ is the volume as a function of time $\tau$. It is based on the QPM EOS of QGP and the isentropic evolution of QGP~\cite{pks1} and given by,
\begin{equation}
 V(\tau, \; b) = v_{0}(b)\left(\frac{\tau_{0}}{\tau}\right)^{\left(\frac{1}{R} -
1\right)},
\end{equation}
where, $R$ is the Reynolds number~\cite{pks1} and $v_{0}(b)$ is the initial
volume at time $\tau_{0}$, given as, $v_{0}(b) = \pi\;(r_{t} -
b/2)^{2}\tau_{0}$, here $r_{t}$ is the radius of fireball created.

Here we use a cooling law for temperature which not only depends on proper time ($\tau$) but also varying with respect to number of participants ($N_{part}$). The cooling law for temperature which connects proper time and $N_{part}$ to the temperature of the system is as follows~\cite{ganesh} :
\begin{equation}
T(\tau) = T_{c}\left(\frac{N_{part}(bin)}{N_{part}(bin_{0})}\right)^{1/3}\left(\frac{\tau_{QGP}}{\tau}\right)^{1/3},
\end{equation}
where $N_{part}(bin_{0})$ is the number of participant corresponding to the most central bin as used in our calculation and $N_{part}(bin)$ is the number of participant corresponding to the bin at which we want to calculate the temperature. $\tau_{QGP}$ is the lifetime of QGP.
\subsection{Cold Nuclear Matter Effect}
We have already discussed about shadowing, absorption and cronin effect as the three main nuclear effects on the charmonium production. Nuclear absorption and Cronin effect are not included in our calculation. We incorporate shadowing as the only dominant CNM effect in the current work.
\subsubsection{Shadowing Effect}
We have used the EPS09 parametrization to obtain the
shadowing for nuclei, with atomic mass number $A$, momentum fraction $x$, and
scale $\mu$, $S^{i}(A,\;x,\;\mu)$~\cite{vogt,Kje}. The spatial variation of shadowing can be given in terms of shadowing and the nucleon density $\rho_{A}(r, z)$ as follows: 
\begin{equation}
 S_{\rho}^{i} (A, x, \mu, r, z) = 1 + N_{\rho}[S^{i}(A, x, \mu) - 1] 
\frac{\int dz\;\rho_{A}(r, z)}{\int dz\;\rho_{A}(0, z)},
\end{equation}
where $N_{\rho}$ is determined by the following normalization condition:
\cite{ganesh},
\begin{equation}
 \frac{1}{A} \int d^{2} r d z \; \rho_{A}(s)\; S_{\rho}^{i} (A, x, \mu, r, z) =
S^{i}(A,\;x,\;\mu). 
\end{equation}
The suppression factor due to shadowing is  defined as;
\begin{equation}
 S_{sh} = R_{AA}(b) =
\frac{d\sigma_{AA}/dy}{T_{AA}\;d\sigma_{pp}/dy} 
\end{equation}
As mentioned in Ref. \cite{ve}, the color evaporation model gives $\sigma_{AA}$ and $\sigma_{pp}$, as follows:
\begin{eqnarray}
 \sigma_{AA} = \int dz_{1}\;dz_{2}\; d^{2}r\; dx_{1}\; dx_{2}\; \nonumber
[f_{g}^{i}(A,\;x_{1},\;\mu,\;r,\; z_{1}) \\ 
\times f_{g}^{j}(A,\;x_{2}, \; \mu, \;
b-r,\; z_{2})\;\sigma_{gg\rightarrow QQ}(x_{1},\;x_{2},\;\mu)],
\end{eqnarray}

\begin{eqnarray}
 \sigma_{pp} = \int dx_{1}\; dx_{2}\;[f_{g}(p,\; x_{1},\;\mu)\; 
f_{g}(p,\; x_{2},\;\mu)\;\\\nonumber\sigma_{gg\rightarrow QQ}(x_{1},\;x_{2},\;\mu)].
\end{eqnarray}
Here, $x_{1}$ and $x_{2}$ are the momentum fraction of the gluons in the two
nuclei and they are related to the rapidity \cite{ganesh}. The superscripts $i$
and $j$ refer to the projectile and target nuclei, respectively.

The function $f_{g}^{i}(A,\;x,\;\mu,\;r,\; z_{1})$ is determined from the gluon
distribution function for proton $f_{g}(p,\; x,\;\mu)$ by using the following relations:
\begin{itemize}
 \item $f_{g}^{i}(A,\;x_{1},\;\mu,\;r,\; z_{1}) = \rho_{A}(s) S^{i} (A,
x_{1}, \mu, r, z)\;f_{g}(p,\; x_{1},\;\mu)$.
\item $f_{g}^{j}(A,\;x_{2}, \; \mu, \;b-r,\; z_{2}) = \rho_{A}(s) S^{j} (A,
x_{2}, \mu, b-r, z)\;f_{g}(p,\; x_{2},\;\mu)$.
\end{itemize}
The value of the gluon distribution function $f_{g}(p,\; x,\;\mu)$ in a proton (indicated by label $p$) has been estimated by using CTEQ6 \cite{jp}.\\

\subsection{Color Screening}
We treat color screening as independent suppression mechanism acting till formation of charmonium bound states followed by gluonic dissociation along with the collisional damping. Original color screening mechanism~\cite{chu} have been modified by Mishra et al.,~\cite{mishra}. In our present work we have used quasiparticle model (QPM) equation of state (EOS) based model in which pressure profile \cite{pks1} and cooling law of pressure are the main ingredients. The cooling law of pressure is given by\\
\begin{equation}
p(\tau,r) = A + \frac{B}{\tau^q} + \frac{C}{\tau} +
\frac{D}{\tau^{c_s^2}}
\end{equation}
where A = -$c_1$, B = $c_2c_s^2$, C = $\frac{4\eta q}{3(c_s^2 - 1)}$ and $D =
c_3$, here  $c_1$, $c_2$, $c_3$ are constants and have been calculated
\cite{pks1,pks2} using different boundary conditions on energy density and pressure. Determining the pressure profile
at initial time $\tau = \tau_{i}$ and at screening time $\tau = \tau_{s}$ we
get, 
\begin{equation}
p(\tau_i,r) = A + \frac{B}{\tau_i^q} + \frac{C}{\tau_i}
+ \frac{D}{\tau_i^{c_s^2}} = p(\tau_i,0)\,h(r)
\end{equation}

\begin{equation}
p(\tau_s,r) = A + \frac{B}{\tau_s^q} + \frac{C}{\tau_s} +
\frac{D}{\tau_s^{c_s^2}} = p_{QGP},
\end{equation}
here $p_{QGP}$ is the pressure of QGP inside screening region, required to
dissociate $J/\psi$, as determined by QPM EOS of QGP~\cite{pks1,pks2}.
After combining cooling law and pressure profile and equating screening time to the dilated formation time, we determined the radius of screening region $r_{s}$.

Assuming that $c\bar c$ is formed inside screening region at a point whose position vector is $\vec r$. It moves with transverse momentum $p_{T}$ making an azimuthal angle $\phi$ (angle between the transverse momentum and position vector $r_{J/\psi}$). Then the condition for escape of $c\bar c$ without forming charmonium states is expressed as: 
\begin{equation}
\cos\phi \geq Y; \; \; Y = \frac{(r_{s}^{2} - r_{J/\psi}^{2})m -
\tau_{F}^{2}p_{T}^{2}/m}{2r_{J/\psi}\tau_{F}p_{T}},
\end{equation}
where, $r_{J/\psi}$ is the position vector at which the charm, anti charm quark pair is
formed, $\tau_{F}$ is the proper formation time required for the formation of
bound states of $c\bar{c}$ from correlated $c \bar c$ pair and $m$ is
the mass of charmonia ($m = M_{J/\psi},\;\; M_{\chi_{c}},\;\; M_{\psi^{'}}$ for different resonance states of charmonium).
\begin{center}
\subsubsection{Survival Probability}
\end{center}
In the color screening scenario, the survival probability of charmonia in QGP medium can
be expressed as;
\begin{equation}
S_c(p_T,N_{part}) = \frac{2(\alpha + 1)}{\pi R_T^2} \int_0^{R_T} dr\,
r\,\phi_{max}(r) \left \{ 1 - \frac{r^2}{R_T^2} \right \}^\alpha,
\end{equation}
where $\alpha = 0.5$~\cite{chu,mishra}.

The condition on azimuthal angle $\phi_{max}$ given by Eq.(4) is
expressed as;
\begin{center}
 $\phi_{max}(r) =  \left\{ 
\begin{tabular}{c}
\vspace{2mm}
$\pi$  $\;\;$ if $\;\;$   $Y\leq -1$\\
\vspace{2mm}
$\pi - cos^{-1}|Y|$ $\;\;$ if $\;\;$  $0\geq Y \geq -1$\\
\vspace{2mm}
$cos^{-1}|Y|$ $\;\;$ if $\;\;$ $0\leq Y \leq -1$\\
\vspace{2mm}
 $0$ $\;\;$ if $\;\;$ $Y\geq 1$
\end{tabular}
\right\}$
\end{center}

Then we have obtained $p_{T}$ integrated survival probability in the color
screening scenario, given as \cite{pks1,pks2}; 
\begin{equation}
 S_{c}(N_{part}) = \frac{\int_{p_{Tmin}}^{{p_{Tmax}}} S(p_{T}, N_{part}) d p_{T}
}{\int_{p_{Tmin}}^{{p_{Tmax}}} d p_{T} }
\label{sp1}
\end{equation}
 for $J/\psi$, $\chi_{c}$ and $\psi^{'}$ denoted by $S_{c}^{J/\psi}$,
$S_{c}^{\chi_{c}}$ and $S_{c}^{\psi^{'}}$, respectively. The range of $p_T$ values used here is allowed by the corresponding experimental data. 

 
\subsection{Net Survival Probability}
\begin{equation}
 N_{J/\psi}^{i}(\tau_{0},b) = N_{J/\psi}(\tau_{0},b)\;S_{sh}
\label{njpsii}
\end{equation}
In Eq. (\ref{tq1}), we have replaced $N_{J/\psi}(\tau_{0},\;b)$ with
$N_{J/\psi}^{i}(\tau_{0},b)$ from Eq. (\ref{njpsii}) and re-obtained the Eq.
(\ref{tq1}), as follows;
\begin{eqnarray}
 N_{J/\psi}^{f}(\tau_{f},b)=\epsilon(\tau_{f})\left[N_{J/\psi}^{i}(\tau_{0},b) + N_{c\bar{c}}^{2} \int_{\tau_{0}}^{\tau_{f}}
\Gamma_{F,nl}\;[V(\tau) \epsilon(\tau)]^{-1} d\tau \right]
\label{tqf}
\end{eqnarray}
Now the survival probability due to shadowing and gluonic dissociation along with collisional damping can be written as;
\begin{equation}
 S_{g}^{J/\psi} = \frac{N_{J/\psi}^{f}(\tau_{f},b)}{N_{J/\psi}(\tau_{0},b)}
\label{sp2}
\end{equation}
In this model we have assumed that at initial stage of QGP the color screening
would be dominated independent to the gluonic dissociation but at later stage it
would partially affect the $J/\psi$ formation and can be coupled with
gluonic dissociation. So we have incorporated the color screening at the initial
stage of QGP followed by gluonic dissociation along with the collisional damping including
shadowing. We expressed the survival probability using Eq. (\ref{sp1}) and
Eq. (\ref{sp2});
\begin{equation}
 S_{f}^{J/\psi} = S_{c}^{J/\psi}\;S_{g}^{J/\psi}
\end{equation}
In the same way we calculated the survival probability for $\chi_{c}$
and $\psi^{'}$, written as, $S_{f}^{\chi_{c}}$ and $S_{f}^{\psi^{'}}$,
respectively.\\
It has been observed that only $60\%$ of $J/\psi$ come up by direct
production whereas $30\%$ is from the decay of $\chi_{c}$ and $10\%$ is form
the decay of $\psi^{'}$, so the net survival probability $S_{J/\psi}$ of
a mixed system after incorporating feed-down is given as:
\begin{equation}
 S_{J/\psi} = \frac{0.60\;N_{J/\psi}\;S_{f}^{J/\psi} + 0.30\;N_{\chi_{c}}\;S_{f}^{\chi_{c}} +
0.10\;N_{\psi^{'}}\;S_{f}^{\psi^{'}}}{0.60\;N_{J/\psi}+0.30\;N_{\chi_{c}}+0.10\;N_{\psi^{'}}}
\end{equation}

\section{Results and Discussions}
Before going to our results, we want to state that some short-hand notations have been used by us to show the different physical processes in the plots. We have used GD for gluonic dissociation, CD for collisional damping, CS for colour screening, S for shadowing as a CNM effect, FD for the feed-down correction and R for the regeneration via recombination of $c$ and $\bar{c}$ quarks.\\
\begin{figure}[h!]
\includegraphics[scale = 0.30]{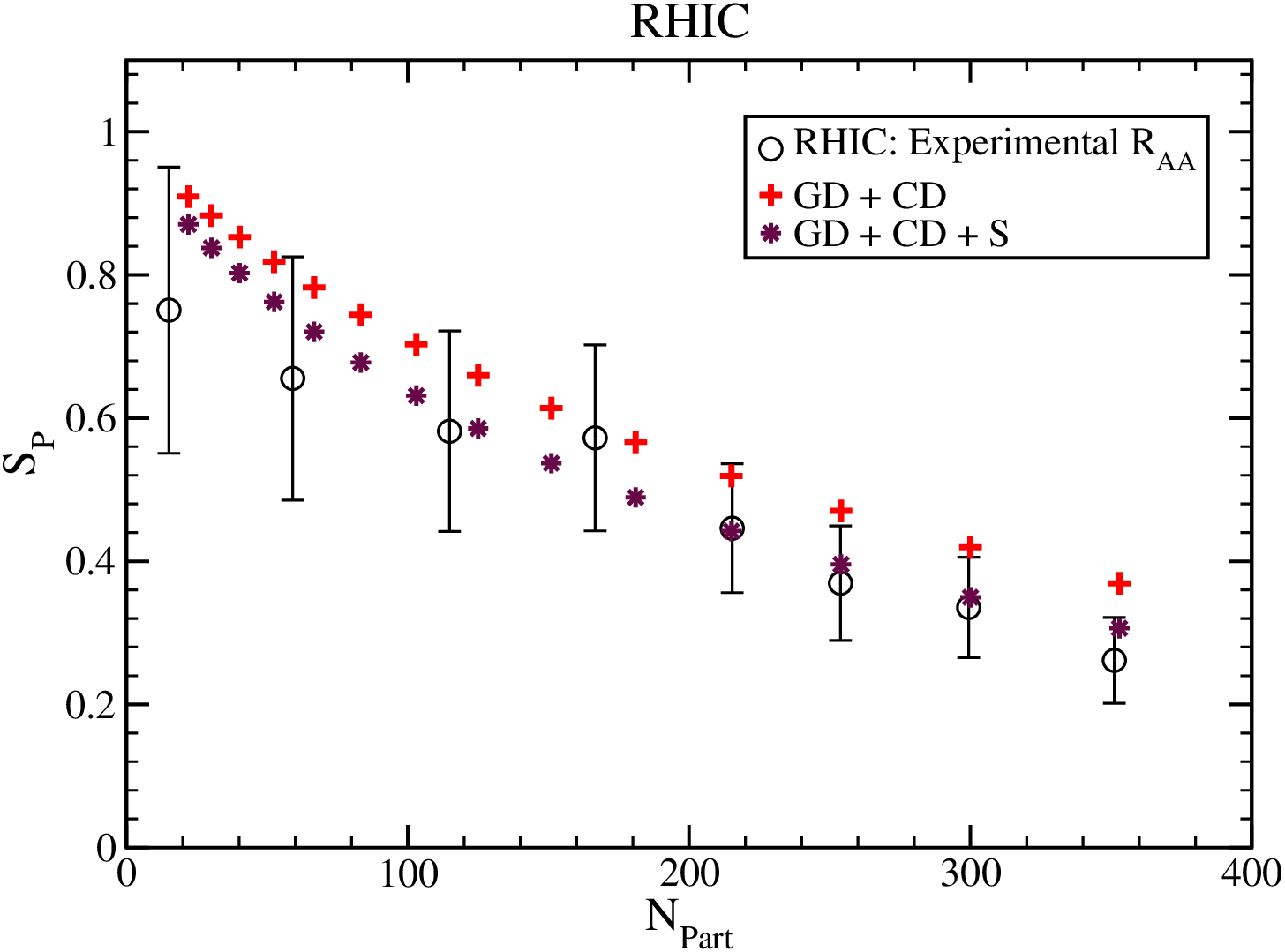}
\caption{(Color online) Survival probability ($S_p$) of $J/\psi$ in QGP medium versus $N_{part}$ including gluonic dissociation (GD) and collisional damping (CD) at RHIC energy with and without shadowing effect (S).}
\end{figure}

\begin{figure}[h!]
\includegraphics[scale=0.30]{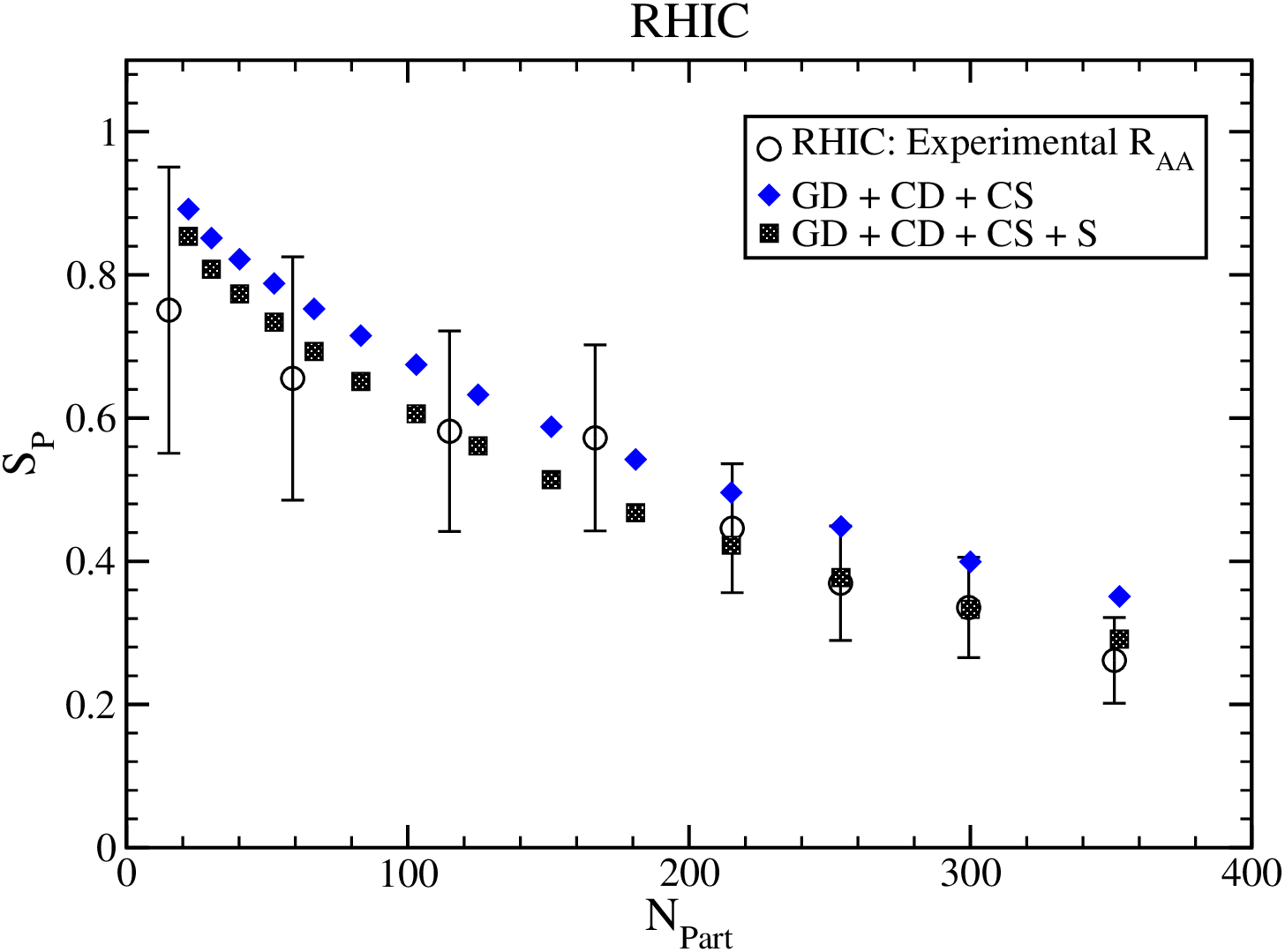}
\caption{(Color online) Survival probability versus $N_{part}$ with gluonic dissociation (GD) and collisional damping (CD) along with color screening (CS) at RHIC with and without shadowing effect (S).}
\end{figure}

\begin{figure}[h!]
\includegraphics[scale=0.30]{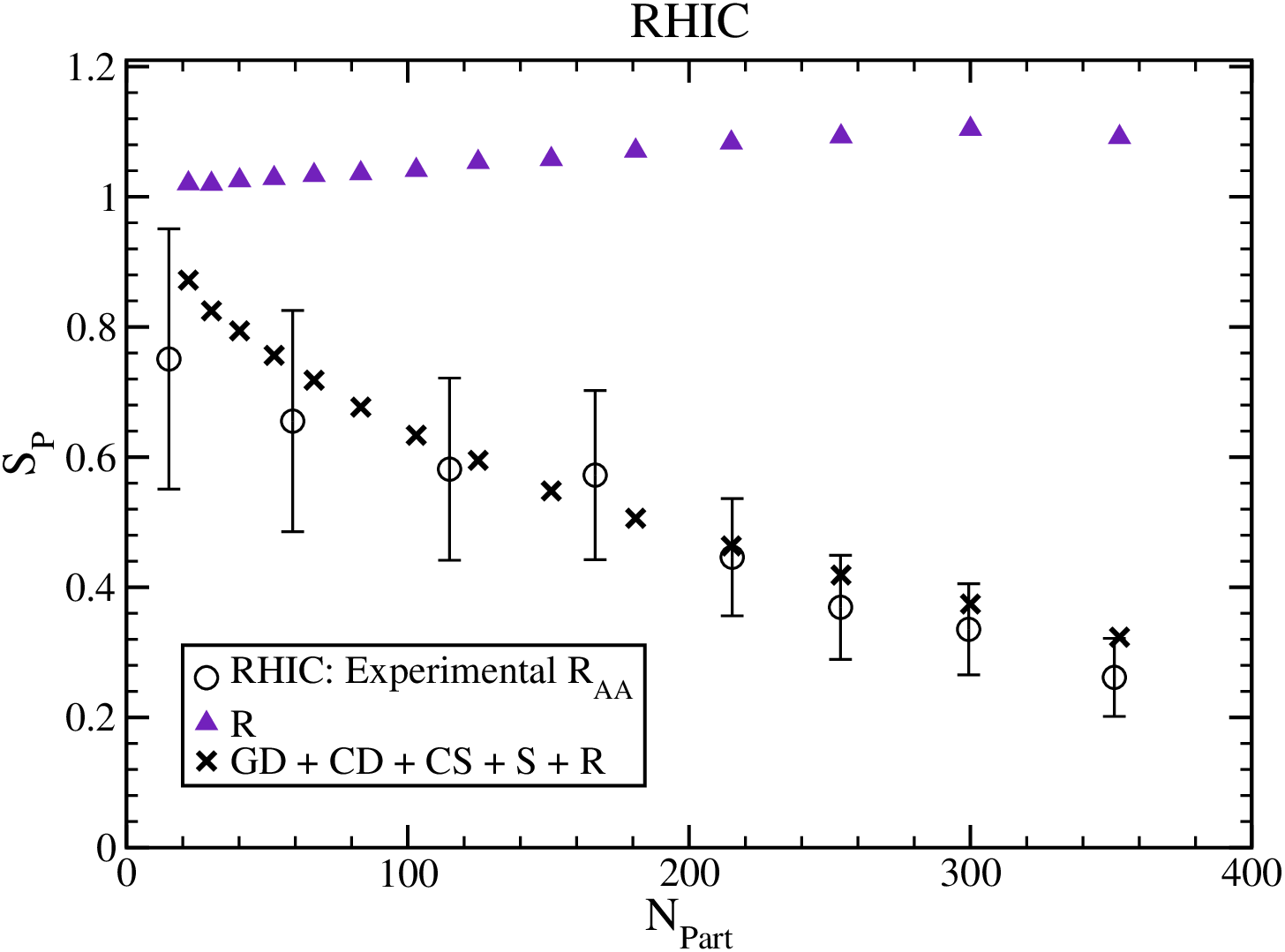}
\caption{(Color online) Same as Fig. 6 but recombination effect (R) is also included. Only recombination (R) is also shown here.}
\vskip 0.03in
\end{figure}

\begin{figure}[h!]
\includegraphics[scale=0.30]{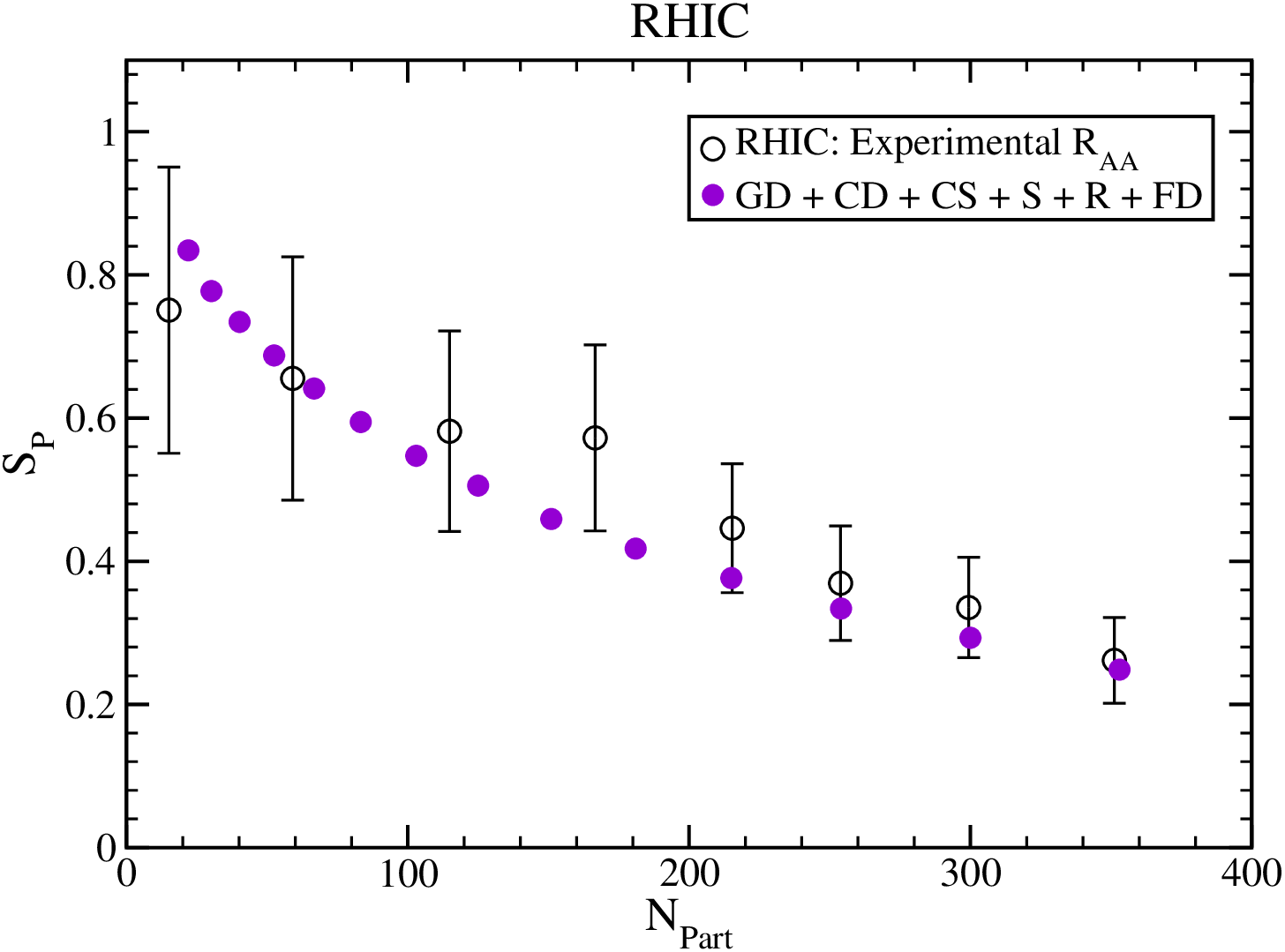}
\caption{(Color online) Same as Fig. 7 but feed-down (FD) due to higher resonances namely, $\chi_c$ and $\psi^{'}$ is also included.}
\end{figure}

Fig. 4 depicts our prediction on suppression (in terms of $p_T$ integrated survival probability, $S_p$) of $J/\psi$ in the QGP medium formed at RHIC as a function of $N_{part}$ at mid rapidity arising due to gluonic dissociation and collisional damping with and without shadowing as a CNM effect. The experimental data on $J/\psi$ suppression obtained from the PHENIX experiment at RHIC~\cite{adare} at center of mass energy, $\sqrt{s_{NN}}=200$ GeV are also shown on the same graph for comparison with our predicted results. It is obvious from the above figure that our result under-predicts the observed $J/\psi$ suppression data without including shadowing. However, after including shadowing, it captures the experimental data on suppression reasonably well. In Fig. 5, we have plotted survival probability versus $N_{part}$ after including color screening along with the gluonic dissociation and collisional damping with and without shadowing effect. Our predicted result is almost similar to the shown in Fig. 4 since color screening only affects on higher charmonium states at RHIC energy. In Fig. 6, we have plotted variation of recombination factor at RHIC energy versus $N_{part}$. This graph indicates that at RHIC energy, recombination for $J/\psi$ increases very slowly with $N_{part}$ and reaches to slightly greater than unity at the most central collisions. All the above mentioned suppression (GD+CD+CS+S) along with the recombination are also plotted on the same graph which obviously will be nearly identical to the variation shown in Fig. 5 since the recombination factor turns out to be small even with higher recombination reactivity at RHIC as compared to LHC. This trend of reactivity is due to the low initial momentum of $c$ and $\bar c$ quarks (in contrary to that happens at LHC due to high initial momentum) which participate in the secondary charmonia formation at RHIC energy. Furthermore, the recombination is mainly governed by the number of ${c\bar c}$ pair produced initially and reactivity of the uncorrelated $c$ and $\bar c$ pair. Due to less number of ${c\bar c}$ pair produced during the initial stage of collisions at RHIC energy and its quadratic dependence on $N_{c\bar c}$, the recombination is found to be small even with significant recombination reactivity at RHIC energy. So far, we have not included feed-down due to the decay of higher resonances of charmonium namely, $\chi_c$ and $\psi^{'}$ to $J/\psi$. Therefore, survival probability of $J/\psi$ with contributions coming from all the suppression mechanisms (GD+CD+CS+S) mentioned above along with the recombination and feed-down via the decay of $\chi_c$ and $\psi^{'}$ to $J/\psi$ is plotted with respect to $N_{part}$ in Fig. 7. After including the feed-down due to higher charmonium reasonances, $J/\psi$ suppression increases at all centralities at RHIC energy but still within experimental error bars. Thus, our results agree reasonably well with the suppression data at mid rapidity obtained from the PHENIX experiment at RHIC at center of mass energy, $\sqrt{s_{NN}}=200$ GeV. 
\begin{figure}[ht]
\includegraphics[scale=0.30]{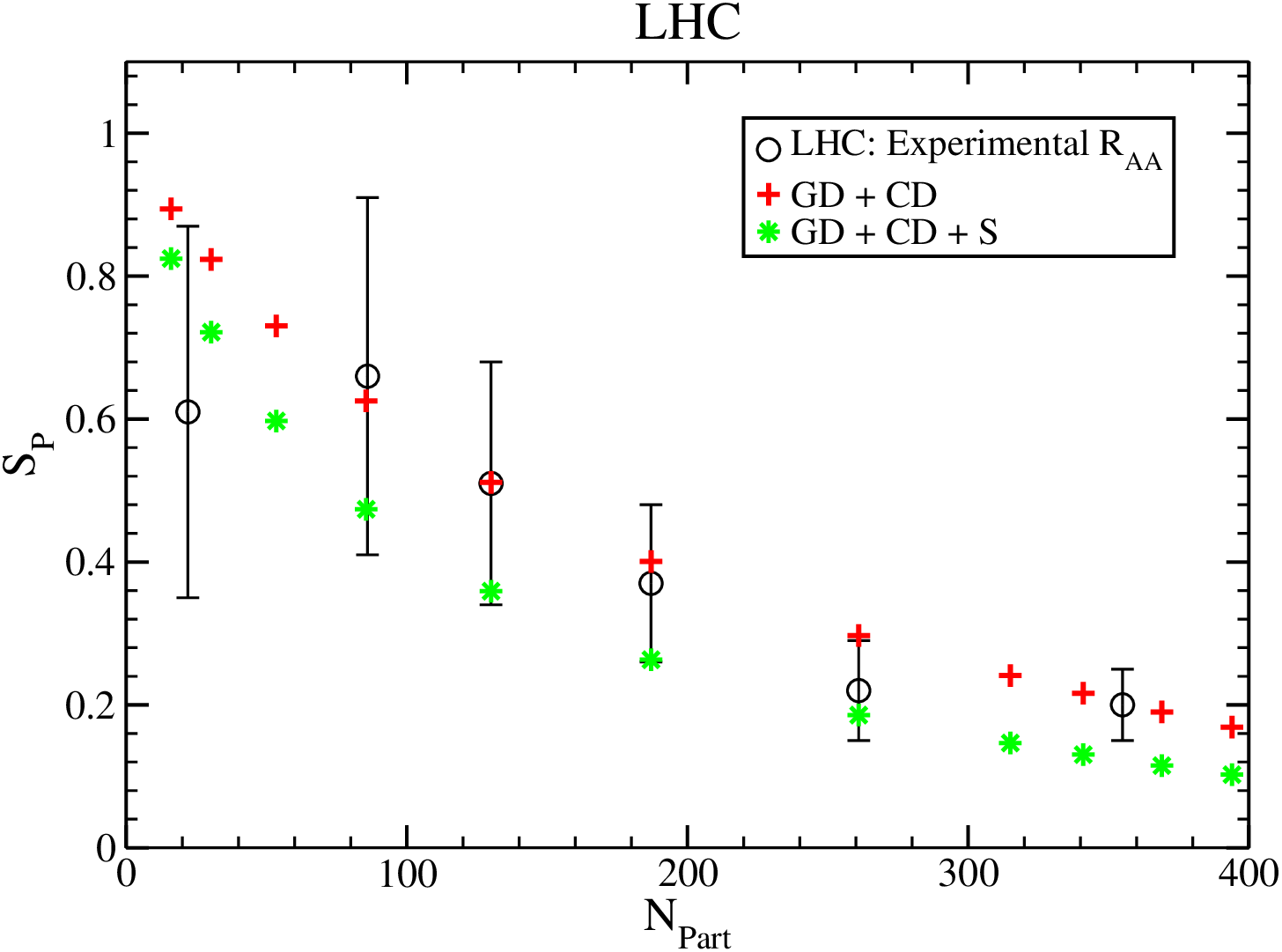}
\caption{(Color online) Survival probability ($S_p$) of $J/\psi$ in QGP medium versus $N_{part}$ including gluonic dissociation (GD) and collisional damping (CD) at LHC energy with and without shadowing effect (S).}
\end{figure}

\begin{figure}[ht]
\includegraphics[scale=0.30]{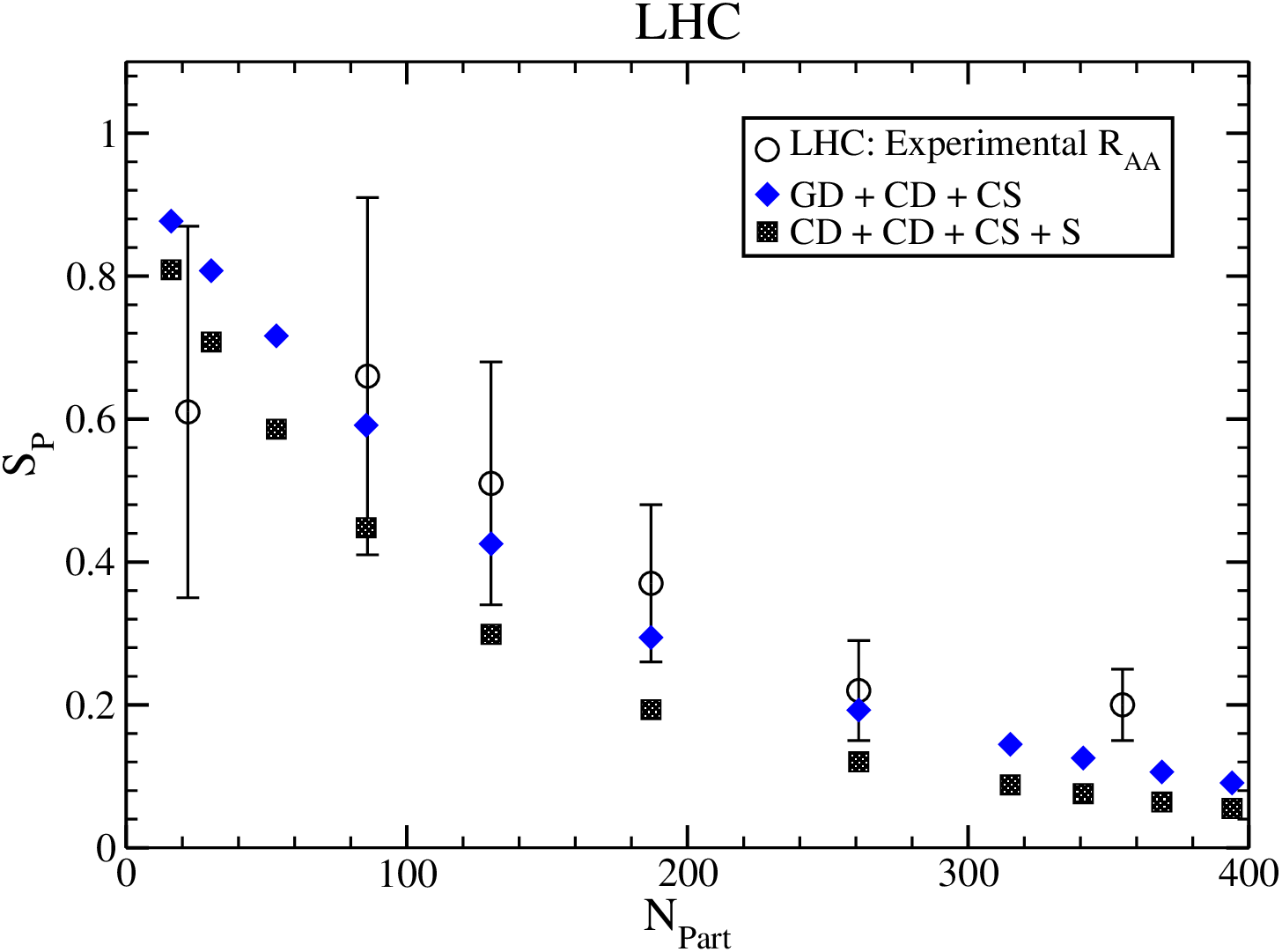}
\caption{(Color online) Survival probability versus $N_{part}$ with gluonic dissociation (GD) and collisional damping (CD) along with color screening (CS) at LHC with and without shadowing effect (S).}
\end{figure}

\begin{figure}[ht]
\includegraphics[scale=0.30]{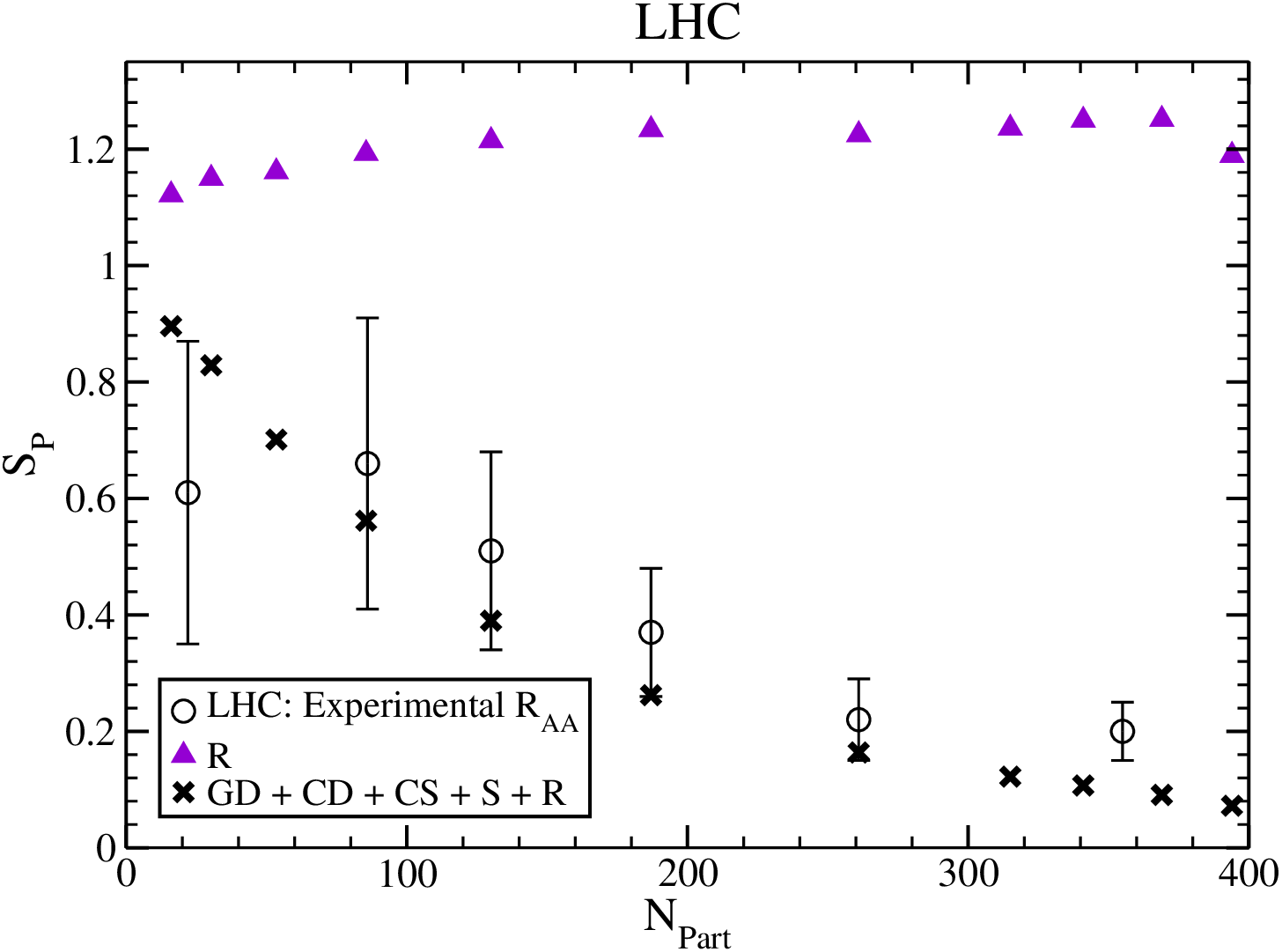}
\caption{(Color online) Same as Fig. 9 but recombination effect (R) is also included. Only recombination (R) is also shown here.}
\vskip 0.03in
\end{figure}

\begin{figure}[ht]
\includegraphics[scale=0.30]{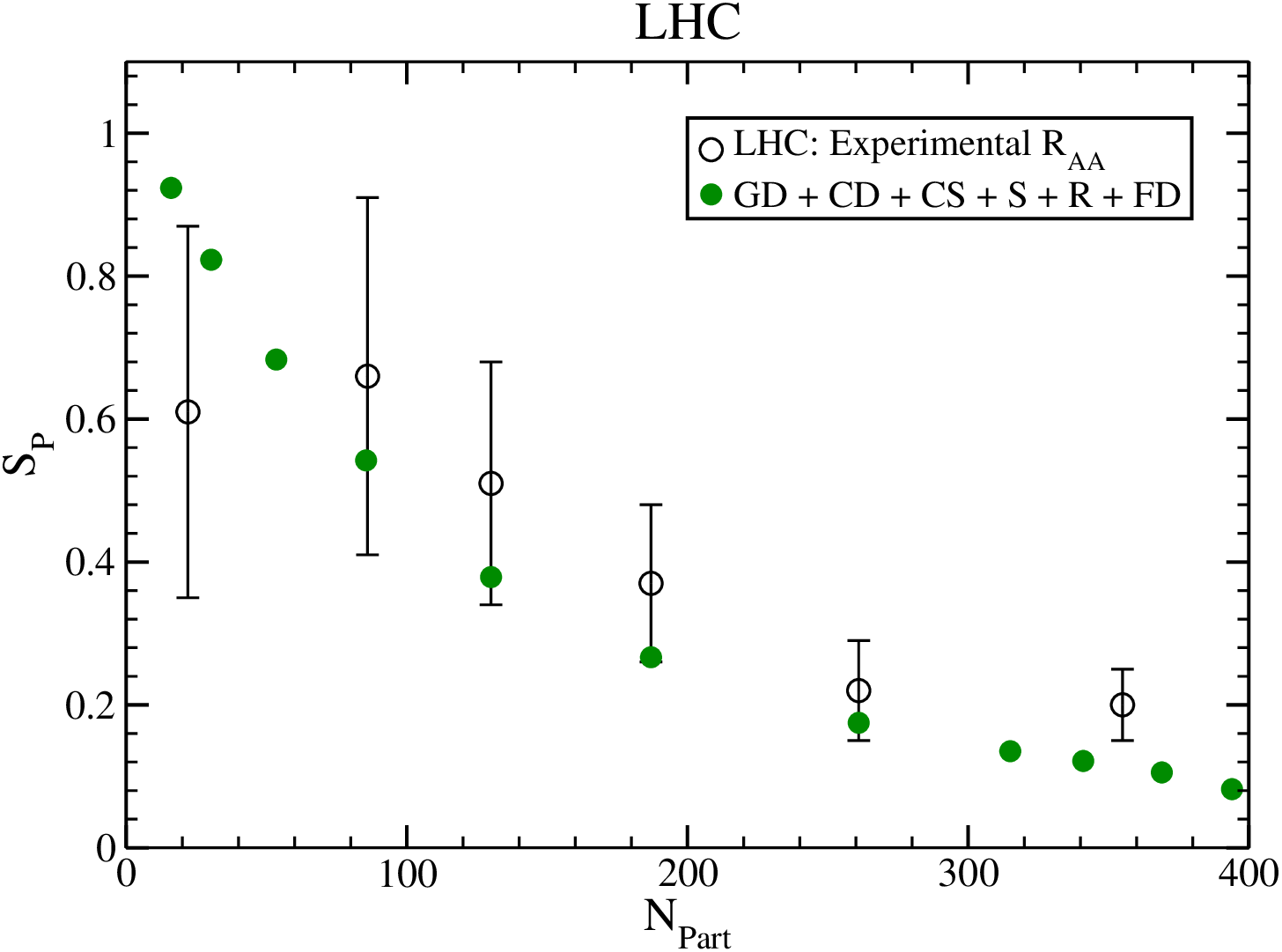}
\caption{(Color online) Same as Fig. 10 but feed-down (FD) due to higher resonances namely, $\chi_c$ and $\psi^{'}$ is also included.}
\end{figure}

Fig. 8 shows $J/\psi$ suppression at mid rapidity versus $N_{part}$ due to the gluonic dissociation and collisional damping with and without shadowing as a CNM effect. The experimental data on $J/\psi$ suppression at mid rapidity obtained from the CMS experiment at LHC~\cite{cms} center of mass energy, $\sqrt{s_{NN}}=2.76$ TeV are also depicted on the same graph for comparison. Our result without shadowing shows good agreement with the experimental data while with shadowing effect, it over predicts the suppression. Fig. 9 includes color screening with the above suppression contributions namely, gluonic dissociation and collisional damping with and without shadowing effect. This figure implies that our predicted result captures data without inclusion of shadowing effect. While with shadowing effect, it shows over suppression. Regeneration factor due to recombination of uncorrelated $c$ and $\bar{c}$ pair at LHC energy and feed-down via decay of higher resonances are two important phenomenon which need to be incorporated to explain the data. Therefore, in the next two figures, we have plotted above recombination effect with all the above mentioned suppression mechanisms (GD+CD+CS+S) with and without feed-down arising due to higher resonance states of the charmonium. Fig. 10 presents all the above suppression contributions and regeneration factor without including feed-down effect due to higher resonances. Only recombination is also presented on the same plot. This figure indicates that recombination of uncorrelated $c\bar c$ pairs at LHC energy varies from around $1.1$ at the most peripheral to around $1.3$ at the most central collisions. Compratively larger recombination occurs here due to the sizable number of $c\bar c$ pairs produced at the LHC even with lower recombination reactivity. This figure clearly indicates that our predicted results capture the trend of data spanning over almost the whole range of $N_{part}$. However, it depicts slight over suppression almost at all centralities.\\

From Fig. 10, it is also evident that at LHC energy, recombination begins to decrease beyond a certain value of $N_{part}$ in contrary to the expectation based on its quadratic dependence on $N_{c\bar c}$ pairs and on volume $V(\tau)$. In fact, recombination has somewhat complex dependence since it does not only depend on the $N_{c\bar c}$ and $V(\tau)$ but also on the recombination reactivity $\Gamma_F$ of uncorrelated $c$ and $\bar c$ which further depends on the temperature of the medium and momentum of $c$ and $\bar c$. Decrease of recombination factor at the highest $N_{part}$ i.e., at the most central collision may be due to the peak value of $\Gamma_F$ occurring at compratively lower temperature (than at RHIC energy) and significantly small value at the highest $N_{part}$. The temperature corresponding to that $N_{part}$ at LHC is too large and at that much higher temperature, the value of $\Gamma_F$ becomes very small (as can be seen from Fig. 3). That is why even with the highest value of $N_{c\bar c}$ at the most central collisions at LHC energy, the small and the reducing trend of $\Gamma_F$ with temperature gives the overall decreasing behavior to the recombination factor. One can observe from Fig. 6 that recombination factor at RHIC does not depict this kind of diminishing trend at the most central collisions. This is because the corresponding $\Gamma_F$ value remains significantly large at RHIC (see Fig. 3) even at the most central collisions (at high temperatures). Thus there is no reducing trend in the recombination factor at the most central collisions at RHIC unlike at LHC energy. Despite of the above facts, recombination factor at LHC is always higher than the corresponding value at RHIC energy for each centrality class (comparison between Figs. 6 and 10) which shows the dominance of $N_{c \bar{c}}^2$ dependence.

  Fig. 11 depicts our results for survival probability of $J/\psi$ with respect to centrality after including contributions from all the suppression mechanisms (GD+CD+CS+S) as a function of $N_{part}$ along with the recombination and feed-down due to decay of higher charmonium resonance states to $J/\psi$. Comparison of Figs. 10 and 11 shows that feed-down due to the decay of higher charmonium states to $J/\psi$ increases the suppression a little bit at LHC energy and slightly over predicts the suppression. However, it still depicts reasonable agreement with the data at LHC energy under the experimental uncertainties. Thus, our analysis shows that the current unified model approach based on the combination of commonly employed suppression and recombination effects present reasonably good agreement with the experimental data at RHIC and LHC energy over the whole range of centrality.

\section{Conclusions}
In conclusions, we have explained recent $J/\psi$ suppression data at mid rapidity obtained from RHIC and LHC experiments using a common formulation based on Debye color screening, gluonic dissociation and collisional damping (popularly termed as Landau damping in the literatures) along with the recombination of uncorrelated $c$ and $\bar c$ pairs in the later stage QGP formation. Shadowing of parton distribution as a CNM effect determined by Vogt approach and feed-down due to decay of higher resonances of charmonium have also been included in the current work. Recombination effect is incorporated via a transport approach using two self coupled transport equations. The QGP medium is assumed to be expanding under Bjorken's scaling law at mid rapidity. Our current model based on the combination of color screening, gluonic dissociation along with collisional damping plus recombination effect explain $J/\psi$ suppression data obtained from both the energies reasonably well without introducing any extra parameter while moving from one set of data at one energy to another set of data at other energy. The explanation of two different sets of suppression data at two different energies, differing by few orders of magnitude, employing a single set of mechanisms without including any new parameter, obviously add importance to our current approach. It is also clear from our results that $J/\psi$ suppression in QGP medium is a result of many complex suppression mechanisms contrary to a single mechanism. Recombination is also important to include to explain charmonium suppression especially at LHC energy. It is worthwhile to note here that although there are few parameters in our current formulation yet not even a single parameter is varied freely in order to explain the suppression data. We have taken all parameter values based on earlier works.         

\noindent
\section{Acknowledgments}
PKS is thankful to Council of Scientific and Industrial Research, Government of India for financial assistance. One of the authors (S. Ganesh) acknowledges Broadcom India Research Pvt. Ltd. for allowing the use of its computational resources required for this work. Captain R. Singh and M. Mishra is grateful to the Department of Science and Technology (DST), New Delhi for financial assistance.

\newpage

\begin{thebibliography}{99}
\bibitem{QWG} N. Brambilla et al., Eur. Phys. J. C {\bf 51}, 1534 (2011).
\bibitem{matsui} T. Matsui and H. Satz, Phys. Lett. B {\bf 178}, 416 (1986).
\bibitem{abreu} M.C. Abreu et al. (NA50 Collaboration, Phys. Lett. B {\bf 477}, 28 (2000); B. Alessandro et al., (NA50 Collaboration), Eur. Phys. J. C {\bf39}, 335 (2005).
\bibitem{arnaldi} R. Arnaldi et al. (NA60 Collaboration), Phys. Rev. Lett.{\bf 99}, 132302 (2007).
\bibitem{adare} A. Adare et al.(PHENIX Collaboration), Phys. Rev. Lett. {\bf 98}, 232301 (2007); arXiv:1201.2251v1 [nucl-exp].
\bibitem{cms} The CMS Collaboration, J. High Energy Phys. {\bf 05}, 063 (2012).
\bibitem{alice} B. Abelev et al. (ALICE Collaboration), Phys. Rev. Lett. {\bf 109}, 072301 (2012).
\bibitem{phenix} A. Adare et al. (PHENIX Collaboration), Phys. Rev. Lett. {\bf 112}, 252301 (2014).
\bibitem{laine} M. Laine, O. Philipsen, P. Romatschke, and M. Tassler, J. High Energy Phys. {\bf 03} (2007) 054.
\bibitem{lata} Lata Thakur, U. Kakade and B. K. Patra, Phys. Rev. D {\bf 89}, 094020 (2014).
\bibitem{brambilla} N. Brambilla, J. Ghiglieri, A. Vairo, P. Petreczky, Phys. Rev. D {\bf 78}, 014017 (2008).
\bibitem{wong} C. Y. Wong, Phys. Rev. C {\bf 72}, 034906 (2005); Y. Park, K. I. Kim, T. Song, S. H. Lee, C. Y. Wong, Phys. Rev. C {\bf 76}, 044907 (2007).
\bibitem{peskin} M. E. Peskin, Nucl. Phys. B {\bf 156}, 365 (1979).
\bibitem{bhanot} G.Bhanot and M. E. Peskin, Nucl. Phys. B {\bf 156}, 391 (1979).
\bibitem{bodwin1} G. T. Bodwin, arXiv:1208.5506v3 [hep-ph].
\bibitem{satz} H. Satz, Acta Phys. Pol. B Supplement {\bf 7}, 49 (2014).
\bibitem{bodwin2} G. T. Bodwin, E. Braaten and G. P. Lepage, Phys. Rev. D {\bf 51}, 1125 (1995) [Erratum-ibid.D55, 5853 (1997)].
\bibitem{bodwin3} G. T. Bodwin, arXiv:1012.4215v1 [hep-ph].
\bibitem{kang1} Z.-B. Kang, J.-W. Qiu and G. Sterman, Phys. Rev. Lett. {\bf 108}, 102002 (2012).
\bibitem{kang2} Z.-B. Kang, Y.-Q. Ma, J.-W. Qiu and G. Sterman, Phys. Rev. D {\bf 90}, 034006 (2014).
\bibitem{baranov1} S. P. Baranov, Phys. Rev. D {\bf 66}, 114003 (2002).
\bibitem{baranov2} S. P. Baranov, A. Szczurek, Phys. Rev. D {\bf 77}, 054016 (2008).
\bibitem{zhou} K. Zhou, N. Xu, Z. Xu and P. Zhuang, Phys. Rev. C {\bf 89}, 054911 (2014).
\bibitem{muller} A. H. Muller and J. W. Qin, Nucl. Phys. B {\bf 268}, 427 (1986).
\bibitem{cronin} J. W. Cronin et al., Phys. Rev. D {\bf 11}, 3105 (1975).
\bibitem{hufner} J. Hufner, Y. Kurihara and H. J. Pirner, Phys. Lett. B {\bf 215}, 218 (1988).
\bibitem{gerschel} C. Gerschel and J. Hufner, Phys. Lett. B {\bf 207}, 253 (1988).
\bibitem{lourenco} C. Lourenco, R. Vogt, H. K. Wohri, J. High Energy Phys. {\bf 02}, 014 (2009).
\bibitem{pbm} P. Braun-Munzinger, J. Stachel, Phys. Lett. B {\bf 490}, 196 (2000); Nucl. Phys. A {\bf 690}, 119c (2001).
\bibitem{andronic} A. Andronic, P. Braun-Munzinger, K. Redlich, J. Stachel, Phys. Lett. B {\bf 571}, 36 (2003); 
Phys. Lett. B {\bf 652}, 259 (2007).
\bibitem{rapp1} L. Grandchamp, R. Rapp, Nucl. Phys. A {\bf 709}, 415 (2002).
\bibitem{rapp2} L. Grandchamp, R. Rapp and G. E. Brown, Phys. Rev. Lett. {\bf 92}, 212301 (2004).
\bibitem{thews1} R. L. Thews, M. Schroedter, J. Rafelski, Phys. Rev. C {\bf 63}, 054905 (2001).
\bibitem{thews2} R. L. Thews, Eur. Phys. J. C {\bf 43}, 97 (2005); Nucl. Phys. A {\bf 702}, 341c (2002).
\bibitem{thews3} R. L. Thews, M. L. Mangano, Phys. Rev. C {\bf 73}, 014904 (2006).
\bibitem{zhang} B. Zhang, C. M. Ko, B. A. Li, Z. W. Lin, S. Pal, Phys. Rev. C {\bf 65}, 054909 (2002).
\bibitem{cassing} E. L. Bratkovskaya, W. Cassing, H. Stocker, Phys. Rev. C {\bf 67}, 054905 (2003).
\bibitem{nendzig} F. Nendzig and G. Wolschin, Phys. Rev. C {\bf 87}, 024911 (2013).
\bibitem{ganesh} S. Ganesh and M. Mishra, Phys. Rev. C {\bf 88}, 044908 (2013); Phys. Rev. C {\bf 91}, 034901 (2015).
\bibitem{pks1} P. K. Srivastava, M. Mishra, C. P. Singh, Phys. Rev. C {\bf 87}, 034903 (2013).
\bibitem{pks2} P. K. Srivastava, S. K. Tiwari, C. P. Singh, Phys. Rev. C {\bf 88}, 044902 (2013).
\bibitem{helmut} H. Satz, J. Phys. G {\bf 32}, 25(R) (2006).
\bibitem{dks} D. Pal, A. Sen, M. G. Mustafa and D. K. Srivastava, Phys. Rev. C {\bf 65}, 034901 (2002)
\bibitem{cern} The ATLAS Collaboration, CERN-PH-EP-2014-172.
\bibitem{ts} Taesoo Song, Kyong Chol Han and Che Ming Ko, arXiv: 1103.6197v2 [nucl-th] (2011).
\bibitem{vogt} R. Vogt, Phys. Rev. C {\bf 81}, 044903 (2010).
\bibitem{Kje} K. J. Eskola, H. Paukkunen, and C. A. Salgado, J. High Energy Phys. {\bf 04}, 065 (2009).
\bibitem{ve} V. Emelyanov, A. Khodinov, S. R. Klein, and R. Vogt, Phys. Rev. Lett. {\bf 81}, 1801 (1998) 
\bibitem{jp} J. Pumplin, D. R. Stump, J. Huston, H. L. Lai, P. M. Nadolsky, and W. K. Tung, J. High Energy Phys. {\bf 07}, 012 (2002).
\bibitem{chu} M.-C. Chu and T. Matsui, Phys. Rev. D {\bf 37}, 1851 (1988).
\bibitem{mishra} M. Mishra, C. P. Singh, V. J. Menon, R. K. Dubey, Phys. Lett. B {\bf 656}, 45 (2007).
 \end{thebibliography}
\end{document}